\documentclass[12pt,a4paper]{amsart}
\usepackage{empheq}
\usepackage{tcolorbox}
\usepackage{framed}
\usepackage{amsmath}
\usepackage[latin9]{inputenc}
\usepackage{amstext}
\usepackage{amssymb}
\usepackage{caption}
\usepackage{longtable}
\usepackage{lscape}
\usepackage{tabularx}
\usepackage{varioref}
\usepackage{amsthm}
\usepackage{graphicx}
\usepackage{txfonts}
\usepackage{pxfonts}
\usepackage{marginnote}
\usepackage{epic}
\usepackage{eepic}
\usepackage{float}
\usepackage{rotating}
\usepackage{epsfig}
\usepackage{indentfirst}
\usepackage{array}
\usepackage{varioref}
\usepackage{tikz}
\usepackage{marginnote}
\usepackage{pgf}
\usepackage{bbm}
\usepackage{appendix}
\usepackage{amsbsy}
\usepackage{latexsym}
\usepackage{amsfonts}
\usepackage{pstricks}
\usepackage{color}
\usepackage[numbers,square,comma, compress]{natbib}
\usepackage{eurosym}
\usepackage{tikz}
\usepackage{pdfpages}
\usepackage{wasysym}
\usepackage{caption}
\usepackage{subcaption}
\usepackage{braket}
\usepackage{simplewick}
\usepackage{yfonts}
\usetikzlibrary{backgrounds}

\usepackage[colorlinks=true,linkcolor=black,citecolor=black,urlcolor=black]{hyperref}

\def\eq{{\epsilon_1}}
\newcommand{\be}{\begin{equation}}
\newcommand{\ee}{\end{equation}}
\newcommand{\ba}{\begin{aligned}}
\newcommand{\ea}{\end{aligned}}

\def\IZ{{\mathbb Z}}

\def\IC{{\mathbb C}}

\def\IP{{\mathbb P}}

\def\({\left(}
\def\){\right)}

\usepackage{
    geometry}
\makeatletter

\pdfpageheight\paperheight
\pdfpagewidth\paperwidth

\numberwithin{equation}{section}
\numberwithin{figure}{section}

\numberwithin{equation}{section}
\numberwithin{figure}{section}

\def\bref\textbf{\ref}

\renewcommand{\d}{\partial}
\def\gS{{g_{\Sigma}}}

\newcommand{\NN}{\mathbbm{N}}

\newcommand{\ZZ}{\mathbbm{Z}}

\newcommand{\et}{\epsilon_2}

\def\ba{\pmb{a}}

\restylefloat{figure}

\usepackage[parfill]{parskip}
\begin{document}
%
\title[]{Blowup Equations and Holomorphic Anomaly Equations}
	
\author[]{Kaiwen Sun
          }
\address{
Korea Institute for Advanced Study, 85 Hoegiro, Dongdaemun-gu, Seoul, South Korea
}

\email{ksun@kias.re.kr
}


\begin{abstract}
Blowup equations and holomorphic anomaly equations are two universal yet completely different approaches to solve refined topological string theory on local Calabi-Yau threefolds corresponding to A- and B-model respectively. The former originated from comparing Nekrasov partition functions of 4d $\mathcal{N}=2$ gauge theories on $\Omega$ defomed spacetime $\IC^2_{\eq,\et}$ and its one-point blown-up, while the latter takes root in the degeneration of wordsheet Riemann surfaces. The relation between the two approaches is an open question. In this short note, we find a novel recursive equation governing their consistency, which we call the consistency equation. This new equation computes the modular anomaly of blowup equations order by order. The consistency equation also suggests a non-holomorphic extension of blowup equations.
\end{abstract}
	
\maketitle
\section{Introduction}
\label{section.01}

\subsection{Blowup equations}
The blowup equations first proposed by Nakajima and Yoshioka as the functional equations of Nekrasov partition functions are now known to exist for 4d $\mathcal{N}=2$ gauge theories \cite{Nakajima:2003pg}, 5d $\mathcal{N}=1$ gauge theories \cite{Nakajima:2005fg,Gottsche:2006bm,Nakajima:2009qjc,Keller:2012da,Kim:2019uqw,KimSun}, 6d $(1,0)$ gauge theories \cite{Gu:2018gmy,Gu:2019dan,Gu:2019pqj,Gu:2020fem} and their untwisted/twisted circle compactifications \cite{Kim:2020hhh}, 6d $(2,0)$ and 5d $\mathcal{N}=1^*$ theories \cite{Gu:2019pqj,Duan:2021ges}, and in general the refined topological string theory on local Calabi-Yau threefolds \cite{Huang:2017mis}. Albeit the striking success in solving the instanton partition functions and refined BPS invariants, many important questions remain unanswered.
In particular, the only proved cases are some 4d and 5d $SU(N)$ gauge theories and related toric Calabi-Yau threefolds \cite{Nakajima:2003pg,Nakajima:2005fg,Gottsche:2006bm,Nakajima:2009qjc}. One of the most imminent questions is perhaps the relation between blowup equations and holomorphic anomaly equations \cite{Bershadsky:1993cx}, which are also known to exist for 4d, 5d, 6d supersymmetric gauge theories and refined topological strings. In this paper, we address this issue in the most general setting of local Calabi-Yau threefolds and refined topological strings.

Recall a local Calabi-Yau threefold $X$ is the anti-canonical line bundle over a complex surface $\mathcal{O}(-S)$. The mirror Calabi-Yau of $X$ can be effectively reduced to a Riemann surface $\Sigma$ of genus $\gS$. It is convenient to
divide all the K\"{a}hler moduli $t_i$ of $X$, $i=1,2,\dots,b$ to true moduli $T_I\,$, $I=1,2,\dots,\gS$ and mass parameters $m_i\,$, $i=1,2,\dots,b-\gS$ corresponding to compact and non-compact 2-cycles of $X$ respectively.
For an arbitrary vector $r\in\IC^b$, we define the $\Lambda$ factor as
\be\label{eq:blowupZ}
\Lambda(t,\eq,\et,r):=	\sum_{ N \in \mathbb{ Z}^{g_{{\Sigma}}}}  (-)^{|N|}\frac{{Z}(t+ \eq R,\eq,\et-\eq)\,{Z}(t+ \et R,\eq-\et,\et  )}{Z(t,\eq,\et)}.
\ee
Here $Z(t,\eq,\et)$ is the full partition function of refined topological strings on Calabi-Yau $X$ and $R:=C\cdot N+r/2$ where $C$ is the intersection matrix between the 2-cycles and the $\gS$ divisors in $\Sigma$.
The main assertion of blowup equations is that there exists vector $r$ such that $\Lambda(t,\eq,\et,r)$ is independent from all true moduli $T$. We call such $r$ vectors as the \emph{$r$ fields}. In the cases $\Lambda\equiv 0$, we call the $r$ fields \emph{vanishing}, otherwise \emph{unity} and $\Lambda$ factor can be written as $\Lambda(m,\eq,\et,r)$. A prerequisite for the $r$ fields is that for all non-vanishing refined BPS invariants $N^{{ d}}_{j_L, j_R} $ of Calabi-Yau $X$ with curve classes ${d}$ and spin $j_{L,R}$,
\be
\label{eq:rcondition}
{2j_L + 2 j_R-1}= {{ r} \cdot {d}}\textrm{ mod }2.
\ee
This is called the \emph{$B$-field condition} or \emph{checkerboard pattern} \cite{Huang:2017mis}. Besides, it is easy to see from blowup equation \eqref{eq:blowupZ} that the $r$ fields are only defined up to equivalence $r\sim r+2C\cdot N$ for $N\in\NN$. Thus the number of vanishing/unity $r$ fields are defined modulo such equivalence.

Recall the free energy $F(t,\epsilon_1,\epsilon_2):=\log Z(t,\epsilon_1,\epsilon_2)$ of refined topological string theory has the following $\eq,\et$ expansion
\be
F(t,\epsilon_1,\epsilon_2)=\sum_{n,g=0}^{\infty}(\epsilon_1+\epsilon_2)^{2n}(\epsilon_1\epsilon_2)^{g-1}{F}_{(n,g)}({t}),\quad\quad n,g\in\IZ.
\ee
For an arbitrary $r$ vector, we can make the following most general expansion of the $\Lambda$ factor in \eqref{eq:blowupZ},
\be\label{LambdaExpansion}
\Lambda(t,\epsilon_1,\epsilon_2,r)=\sum_{n,g=0}^{\infty}(\epsilon_1+\epsilon_2)^{2n}(\epsilon_1\epsilon_2)^{g}{\Lambda}_{(n,g)}({t},r),\quad\quad n\in\IZ/2,g\in\IZ.
\ee
Note the terms with $(\epsilon_1\epsilon_2)^{-1}$ always cancel with each other such that $\Lambda$ has no pole in $\epsilon_1$ and $\epsilon_2$.
Then from \eqref{eq:blowupZ}, we obtain infinite many equations for  ${\Lambda}_{(n,g)}({t},r)$ called 
the \emph{component equations}. For example, the leading order is
\be
\Lambda_{(0,0)}=\sum_{ N \in \mathbb{ Z}^{\gS}} (-)^{|N|} \exp{ \left(-\frac{1}{2} R^2 F_{(0,0)}''+F_{(0,1)}-F_{(1,0)}\right)}.
\ee
Here we use the abbreviation $R^mF^{(m)}_{(n,g)}=\sum_{i_1,\dots,i_m} R_{i_1}\cdots R_{i_{m}}\d_{ t_{i_1}}\cdots\d_{t_{i_{m}}}F_{(n,g)}.$
This equation is actually a generalization of the contact term equations in supersymmetric $SU(N)$ gauge theories, see for example \cite{Losev:1997tp,Lossev:1997bz,Gorsky:1998rp,Marino:1999qk}. Thus we call it \emph{generalized} or \emph{the zeroth contact term equation}.
The subleading order of blowup equation \eqref{eq:blowupZ} is
\be
\Lambda_{(\frac12,0)}=\sum_{ N \in \mathbb{ Z}^{\gS}} (-)^{|N|} \left(-\frac{1}{6} R^3 F_{(0,0)}^{(3)}+R\left(F_{(0,1)}'+F_{(1,0)}'\right)\right)\exp{ \left(-\frac{1}{2} R^2 F_{(0,0)}''+F_{(0,1)}-F_{(1,0)}\right)} .
\ee
We call this component equation as \emph{the first contact term equation}. Both two contact term equations only involve refined free energies $F_{(0,0)}$, $F_{(0,1)}$ and $F_{(1,0)}$.

Topological strings in the B-model are closely related to modular forms. In the seminal paper of Aganagic, Bouchard and Klemm \cite{Aganagic:2006wq}, the topological string amplitudes depending on polarization are realized as either holomorphic quasi-modular forms or almost holomorphic forms of weight zero with respect to the monodromy group $\Gamma$ of the periods of the Calabi-Yau threefolds. In our current setting, for refined topological string on local Calabi-Yau $X$, the refined free energies and their $t$ derivatives $F_{(n,g)}^{(m)}(t)$ are holomorphic quasi-modular forms of weight zero with respect to monodromy group $\Gamma\subset Sp(2\gS,\ZZ)$.

As was demonstrated in \cite{Huang:2017mis}, for any $r$ vector and arbitrary $n,g\ge 0$, the function ${\Lambda}_{(n,g)}({t,r})$ are in general quasi-modular forms of weight zero. The non-triviality of blowup equations lies in that for specially chosen $r$ -- the $r$ fields, ${\Lambda}_{(n,g)}({t,r})$ are not just quasi-modular but in fact \emph{modular invariant}! In physics terms, they contain no modular anomaly: $\delta\Lambda_{(n,g)}=0$. This observation on the $\Lambda$ factors relates blowup equations to holomorphic/modular anomaly equations.

\subsubsection{Remark}
\textit{There exist local Calabi-Yau threefolds that have only vanishing $r$ fields such as massless local half K3 \cite{Huang:2017mis} and many elliptic non-compact Calabi-Yau threefolds associated to 6d (1,0) SCFTs with half-hypers \cite{Gu:2020fem}, or only unity $r$ fields such as resolved conifold \cite{Huang:2017mis} and many geometries associated to 6d (2,0) ADE SCFTs \cite{Gu:2019pqj,Duan:2021ges}.}

\subsection{Refined holomorphic and modular anomaly equations}

The holomorphic anomaly equations were first proposed by Bershadsky, Cecotti, Ooguri and Vafa \cite{Bershadsky:1993cx} for topological string theory on compact Calabi-Yau threefolds. They were later generalized to the refined holomorphic anomaly equations for gauge theories and refined topological strings in \cite{Krefl:2010fm,Huang:2010kf}. In such situation, one consider the full \emph{non-holomorphic refined free energies} $F_{(n,g)}(t,\bar{t})$ rather than their holomorphic limit $F_{(n,g)}(t)$. The non-holomorphic refined free energies are defined as certain integration over the moduli space of Riemann surfaces \cite{Huang:2011qx} and the non-holomorphicity -- holomorphic anomaly takes root in the contribution from the boundary of the moduli space where the Riemann surfaces degenerate. Careful analysis on the degeneration of Riemann surfaces leads to the celebrated BCOV holomorphic anomaly equations and their refined generalization.
We refer to \cite{Klemm:2015iya} for an excellent review on the background for this subsection.

For general local Calabi-Yau threefold $X$,
refined holomorphic anomaly equation for the refined topological string free energies can be written as \cite{Huang:2010kf} 
\begin{eqnarray} \label{RHAE}
\frac{\partial}{\partial {{\bar t}_{\bar \imath}}}F^{(n,g)}&=&\frac{1}{2} \bar{C}^{jk}_{\bar \imath}\Bigg(F^{(n,g-1)}_{j,k}+
\sum_{n',g'}'F^{(n',g')}_{j}F^{(n-n',g-g')}_{k}\Bigg)\ ,\qquad n+g>1.
\end{eqnarray} 
Here the prime on the summation means $(n',g')$ runs from $(0,0)$ to $(n,g)$ but excludes $(0,0)$ and $(n,g)$. $F^{(n,g)}_{i_1,\ldots, i_m}$ is the covariant derivative of  the non-holomorphic refined free energy $D_{i_1}\ldots D_{i_m}F^{(n,g)}(t,\bar{t})$. Besides, $\bar C^{ij}_{{\bar \imath}}=e^{2{\mathcal K}}  G^{j\bar \jmath} G^{k\bar k} C_{\bar \imath\bar \jmath \bar k}$ contains the K\"ahler potential, the metric $G^{j\bar \jmath}$, and the complex conjugate of the Yukawa coupling 
$C_{ijk}=F^{(0,0)}_{i,j,k}$. This equation has clear geometric meaning: the first term in right hand side means the degeneration of a genus $g$ curve to a genus $g-1$ Riemann surface, while the second term means the degeneration of a genus $g$ Riemann surface to two disconnected Riemann surfaces. In the meantime, the $n$ marked points on the Riemann surface are randomly distributed. 

To make contact with blowup equations \eqref{eq:blowupZ} which only involve holomorphic free energy, we need make use of the \emph{modular anomaly equations} \cite{Huang:2009md} which are closely related to the direct integration of holomorphic anomaly equations with respect to the \emph{propagators} $S^{ij}$ \cite{Bershadsky:1993cx}. The propagator allows one to replace the non-holomorphic derivative in \eqref{RHAE} as 
\be
\frac{\partial}{\partial {{\bar t}_{\bar \imath}}}F^{(n,g)}=\bar{C}^{jk}_{\bar \imath}\frac{\partial F^{(n,g)}}{\partial S^{jk}}.
\ee
Then assuming linear independence, the holomorphic anomaly equation \eqref{RHAE} becomes
\begin{eqnarray} \nonumber
\frac{\partial F^{(n,g)}}{\partial S^{jk}}&=&\frac{1}{2} \Bigg(F^{(n,g-1)}_{j,k}+
\sum_{n',g'}'F^{(n',g')}_{j}F^{(n-n',g-g')}_{k}\Bigg),\qquad n+g>1.
\end{eqnarray} 
This provides an efficient approach to solve the full free energy $F^{(n,g)}(t,\bar{t})$ recursively up to some holomorphic ambiguities which can be further fixed by certain boundary conditions \cite{Huang:2010kf,Huang:2011qx,Huang:2013yta}.

For local Calabi-Yau with mirror curve genus $g_\Sigma=1$, it is well-known the propagator is proportional to the almost holomorphic Eisenstein series $\hat{E}_2(\tau,\bar{\tau})$ defined by
\be
\hat{E}_2(\tau,\bar{\tau})=E_2(\tau)-\frac{3}{\pi\mathrm{Im}(\tau)}.
\ee
Here the Eisenstein series $E_2(\tau)$ is a holomorphic quasi-modular form of weight two. Just as the $\hat{E}_2(\tau,\bar{\tau})$ dependence measures the holomorphic anomaly of $F_{(n,g)}(t,\bar{t})$, in the holomorphic limit, the $E_2(\tau)$ dependence measures the modular anomaly of holomorphic free energies $F_{(n,g)}(t)$.
We denote operator $\delta$ as $\partial_{E_2}$ up to some unimportant constant factor.
Then the refined modular anomaly equation for holomorphic refined free energies and their derivatives with respect to the single true K\"{a}hler modulus can be universally written \cite{Huang:2011qx} as
\be\label{MAE}
\delta F_{(n,g)}^{(m)}=\frac{1}{2}\Bigg(F_{(n,g-1)}^{(m+2)}+\sum_{g'+g''=g,n'+n''+n}^{m'+m''=m}\frac{m!}{m'!m''!}F_{(n',g')}^{(m'+1)}F_{(n'',g'')}^{(m''+1)}\Bigg).
\ee
The summation $(n',g')$ runs from $(0,0)$ to $(n,g)$ and for $g'=0$ or $g'=g$ either $n>0$ or $m>3$.
Besides,
\be
\delta F^{(m)}_{(n,g)}=0,\quad\quad\textrm{for   }\quad 3(n+g-1)+m\le 0.
\ee
This is one main starting point for us to check the consistency between blowup equations and holomorphic anomaly equations. For more discussion on the equivalence between holomorphic anomaly equations and modular anomaly equations, see \cite{Huang:2013eja}. Besides, holomorphic anomaly equations in the Nekrasov-Shatashivili limit $(\eq\to 0,\et\to\hbar)$ can be derived from the quantum WKB expansion of the mirror curves \cite{Huang:2012kn}.

For local Calabi-Yau with mirror curve genus $g_\Sigma>1$, we use the abstract higher dimensional analogy\footnote{Strictly speaking, there is no analogue of the almost holomorphic $E_2$ for $\gS= 2$, but only certain almost meromorphic substitiute of it \cite{Klemm:2015iya}.}  to the Eisenstein series $E_2$ denoted as $E^{IJ}$ \cite{Aganagic:2006wq}. The almost holomorphic one is related to the quasi-modular one by $\hat{ E}^{IJ}(\tau,\bar \tau) = E^{IJ}(\tau)+ (( \tau-{\bar \tau})^{-1})^{IJ}$ and $\hat{ E}^{IJ}(\tau,\bar \tau)$ transform as
$$
{\hat E}^{IJ}(\tau, \bar{\tau}) \rightarrow
{(C\tau +D)^{I}}_{K}\;{(C\tau+D)^{J}}_L \;{\hat E}^{KL}(\tau,\bar{\tau})
$$
under monodromy group $\Gamma\subset Sp(2\gS,\ZZ)$. 
It is expected these $E_{IJ}$ forms can be obtained from
\be
E^{IJ}(\tau)=\frac{\partial}{\partial\tau_{IJ} }\log\phi(\tau)
\ee
with $\phi(\tau)$ as certain scalar Siegel cusp form of degree $\gS$. For example, for local Calabi-Yau with mirror curve of $\gS=1$, the cusp form is just the well-known Ramanujan modular form $\Delta$ of weight 12. While for $\gS=2$, $\phi$ is the famous Igusa cusp form of weight 10 \cite{Igusa}. For $\gS=3$, $\phi$ is expected to be the weight 18 cusp form defined by Tsuyumine \cite{Tsuyumine}. In the case of $\gS=2$, the $E^{IJ}(\tau)$ defined in this way has been used to solve refined topological strings on resolved $\IC^3/\IZ_5$ and resolved $\IC^3/\IZ_6$ Calabi-Yau threefolds \cite{Klemm:2015iya}. 

We use $\delta_{IJ}$ as the derivative with respect to $E_{IJ}(\tau)$ to measure the modular anomaly for refined topological string free energy on general local Calabi-Yau.



\subsection{Main result}

The main result of this paper is the following recursive equation on the modular anomaly of $\Lambda_{(n,g)}$ computed by the refined modular anomaly equations: for an arbitrary vector $r$ such that the zeroth and first contact term equations are satisfied, 
\be\label{deltaIJ}
\delta_{IJ}\Lambda_{(n,g)}=\frac{1}{2}C_I^iC_J^j\sum_{i,j}\Bigg(\partial_i\partial_j\Lambda_{(n,g-1)}+\sum'\big(\partial_i F_{(n_1,g_1)}\partial_j\Lambda_{(n_2,g_2)}+\partial_j F_{(n_1,g_1)}\partial_i\Lambda_{(n_2,g_2)}\big)\Bigg).
\ee
Here $2n\in\IZ_{\ge 0}$ and $g\in\IZ_{\ge 0}$. For $n\in \IZ$, 
 the summation is over all non-negative integers $(n_1,g_1,n_2,g_2)$ such that $n_1+n_2=n$ and $g_1+g_2=g$ excluding $(n_1,g_1)=(0,0)$ and $(n_2,g_2)=(0,0)$. For $n\in \IZ+1/2$, $n_2$ are required to be half integers and the summation  excludes $(n_1,g_1)=(0,0)$ and $(n_2,g_2)=(1/2,0)$. Besides, recall $C_I^i$ is the intersection matrix between the 2-cycles of Calabi-Yau $X$ and the divisors in $\Sigma$. 
 It is convenient to write the above equations for all $(n,g)$ together as
\be
\delta_{IJ}\Lambda=\frac{\eq\et}{2}C_I^iC_J^j\sum_{i,j}\Big(\partial_i\partial_j\Lambda+\partial_i\bar{F}\partial_j\Lambda+\partial_j\bar{F}\partial_i\Lambda\Big).
\ee
Here we define $\bar{F}=F-F_{(0,0)}/(\eq\et)$. More simply we can write it as
\be
\delta\Lambda=\frac{\eq\et}{2}\Big(\partial^2\Lambda+\{\partial\bar{F},\partial\Lambda\}\Big).
\ee

The general form of equation \eqref{deltaIJ} is inspired from the $\gS=1$ cases. 
For local Calabi-Yau threefolds with genus one mirror curve which we will mainly focus on, the above recursive equation can be written as
\be\label{CE1}
\delta\Lambda_{(n,g)}=\frac{1}{2}\Lambda_{(n,g-1)}''+\sum_{n_1,g_1}'F_{(n_1,g_1)}'\Lambda_{(n-n_1,g-g_1)}'\ ,\qquad\quad 2n,g\in\IZ_{\ge 0}.
\ee
Here the derivative is taken with respect to the single true K\"{a}hler modulus.

Once the above recursive equation \eqref{deltaIJ} for modular anomaly of $\Lambda_{(n,g)}$ is established, the consistency between blowup equations and refined modular anomaly equations becomes obvious and trivial. On one hand, the defining property of $r$ fields is that for all true moduli $T_I$, $\partial_{T_I}\Lambda_{(n,g)}=0$. Using the fact that the non-compact 2-cycles have $C_I^i$ components all zero, then \eqref{deltaIJ} implies all $\delta_{IJ}\Lambda_{(n,g)}=0$. On the other hand, $\Lambda$ is already expected to be anomaly free due to the independence 
from the true moduli. Note the importance here lies in that \emph{\eqref{deltaIJ} is computed from refined modular anomaly equations}. This makes direct contact between blowup equations and holomorphic/modular anomaly equations and shows their structure are highly compatible. 
Therefore we call \eqref{deltaIJ} as the \emph{consistency equation}.

\subsection{Outline of the paper}
In section \textbf{\ref{sec:localP2}}, we focus on local Calabi-Yau with $\gS=1$ such as local $\IP^2$ and show in great detail how consistency equation \eqref{CE1} emerges. In section \textbf{\ref{sec:nonholo}}, we propose a non-holomorphic extention of blowup equations for refined topological strings. In section \textbf{\ref{sec:open}}, we summarize and discuss some open questions on blowup equations.

\section{Local Calabi-Yau with genus-one mirror curves}\label{sec:localP2}
Local $\mathbb{P}^2$ is the simplest local toric Calabi-Yau threefold with non-trivial 4-cycles. It has a single parameter $t$ and a genus-one mirror curve. The intersection matrix $C=3$. The blowup equations for local $\mathbb{P}^2$ have been found in \cite{Huang:2017mis} which have one vanishing $r$ field $r_v=3$ and two unity $r$ fields $r_u=\pm 1$. The component equations for $\Lambda_{(n,g)}$ give infinite many identities for modular forms on monodromy group $\Gamma(3)$, for example the unity zeroth contact term equation gives the Euler's Pentagonal number theorem \cite{Huang:2017mis}. Here we want to compute the modular anomaly $\delta \Lambda_{(n,g)}$ from refined modular anomaly equations \eqref{MAE}.

The computation in this section can be seen directly for local $\mathbb{P}^2$, but also can be seen for all local Calabi-Yau threefolds with genus-one mirror curves. In the latter cases, all mass-parameter $m$ dependence is implicit and $t$ is for the single true modulus. Most local Calabi-Yau threefolds with genus-one mirror curves can be obtained as the blown-down of local half-K3 Calabi-Yau threefolds. Their blowup equations have been found in \cite{Gu:2019pqj}.

We divide the computation on the modular anomaly of $\Lambda_{(n,g)}$ to several cases. First the even order of unity blowup equations that is $n+g\in Z$, then the odd order of unity blowup equations that is $n+g\in Z+1/2$, and last the vanishing blowup equations.
\subsection{Unity even orders} Recall given one $r$ vector, the shifts $R=CN+r/2$ with $N$ taking over all integers $\IZ$. 
Let us denote
\be
f(n):=\sum_R \Theta(t,R)R^{2n},\qquad 2n\in \mathbb{Z}_{\ge 0},
\ee
where
\be
\Theta(t,R)=(-)^{|N|} \exp{ \left(-\frac{1}{2} R^2 F_{(0,0)}''+F_{(0,1)}-F_{(1,0)}\right)}.
\ee
It is easy to obtain the following recursive relation
\be\label{fnrecur}
-\frac12F_{(0,0)}'''f(n+1)+(F_{(0,1)}'-F_{(1,0)}')f(n)=f'(n).
\ee
With the initial condition $f(0)=\Lambda_{(0,0)}$, the recursion gives all $f(n)$ for $n=1,2,3,\dots$ Remember we assume the generalized contact term equation holds, i.e. $\Lambda_{(0,0)}$ is independent from the true moduli $t$. Thus $\Lambda_{(0,0)}'=0$ and $\delta\Lambda_{(0,0)}=0 $.

Consider $\Lambda_{(1,0)}$ at first. From blowup equation, we have
\be
\Lambda_{(1,0)}=\sum\Theta(R)\Big(3 F_{(0,2)}-2 F_{(1,1)}+F_{(2,0)}-R^2 F''_{(0,1)}+\frac{1}{2}R^2 F''_{(1,0)}+\frac{1}{24} R^4 F_{0}^{(4)}\Big).
\ee
Using recursive relation (\ref{fnrecur}) to replace those $\sum \Theta(t,R) R^{2n}$, we obtain
\begin{multline}
\Lambda_{(1,0)}=\Lambda_{(0,0)}\bigg( 3 F_{(0,2)}-2 F_{(1,1)}+F_{(2,0)}+(2F''_{(0,1)}- F''_{(1,0)})\frac{F_{(1,0)}'-F_{(0,1)}'}{F_0'''}
\\ \phantom{----}+\frac{F_0^{(4)}}{6(F_0''')^2}\Big(-\frac{F_0^{(4)}}{F_0'''}
(F_{(1,0)}'-F_{(0,1)}')+(F_{(1,0)}''-F_{(0,1)}'')+(F_{(1,0)}'-F_{(0,1)}')^2\Big)\bigg).
\end{multline}
By long computation with direct variations, we find 
\be
\delta \Lambda_{(1,0)}=0.
\ee
The relevant variations from refined modular anomaly equations used here are
\be\label{variationEx}
\begin{aligned}
\delta F_{(0,2)}&=\frac{1}{2}\Big(F_{(0,1)}''+F_{(0,1)}'F_{(0,1)}'\Big),\quad
\delta F_{(1,1)}=\frac{1}{2}\Big(F_{(1,0)}''+2F_{(0,1)}'F_{(1,0)}'\Big),\quad
\delta F_{(2,0)}=\frac{1}{2}F_{(1,0)}'F_{(1,0)}'\ ,\\
\delta F_{(0,1)}'&=\frac{1}{2} F_{(0,0)}'''\ ,\qquad
\delta F_{(1,0)}'=0\ ,\qquad
\delta F_{(0,1)}''=\frac{1}{2}\Big(F_{(0,0)}^{(4)}+F_{(0,0)}'''F_{(0,1)}'\Big),\\
\delta F_{(1,0)}''&=\frac{1}{2}F_{(0,0)}'''F_{(1,0)}'\ ,\qquad
\delta F_{(0,0)}^{(4)}=3F_{(0,0)}'''F_{(0,0)}'''.
\end{aligned}
\ee
Similar computation gives $\delta \Lambda_{(0,1)}=0$ where
\be
\begin{aligned}
\Lambda_{(0,1)}=\sum&\Theta(R)\bigg(-F_{(0,2)}-3F_{(2,0)}+\frac12R^2\big( F''_{(0,1)}+(F'_{(0,1)}+ F'_{(1,0)})^2\big)\\
&\qquad-\frac{1}{6} R^4 F_{0}^{(3)}(F'_{(0,1)}+ F'_{(1,0)})-\frac{1}{24} R^4 F_{0}^{(4)}+\frac{1}{72} R^6 (F_{0}^{(3)})^2\bigg).
\end{aligned}
\ee

For total genus $n+g> 1$, the anomaly of $\Lambda$ factors becomes complicated. However we still manage to find a closed formula for them. Take $\Lambda_{(0,2)}$ for example. From the expansion of blowup equations we find
\be\nonumber
\begin{aligned}
\Lambda_{(0,2)}=\sum&\Theta(R)\Bigg(\left(7  F_{(0,3)} -4  F_{(1,2)} +2  F_{(2,1)} - F_{(3,0)}+\frac{1}{2} (3  F_{(0,2)} -2  F_{(1,1)} + F_{(2,0)} )^2 \right)\\
&-R^2 \left(2  F_{(0,2)}''- F_{(1,1)}'' +\frac{1}{2}  F_{(2,0)}'' +\frac{1}{2} (3  F_{(0,2)} -2  F_{(1,1)} + F_{(2,0)} ) \left(2  F_{(0,1)}'' - F_{(1,0)}'' \right) \right)\\
&+\frac{1}{24} R^4 \left(3 \left( F_{(1,0)}'' -2  F_{(0,1)}'' \right)^2+(3  F_{(0,2)} -2  F_{(1,1)} + F_{(2,0)} )  F_{(0,0)}^{(4)} +2  F_{(0,1)}^{(4)} - F_{(1,0)}^{(4)} \right)\\
&+R^6 \left(\frac{1}{48}  F_{(0,0)}^{(4)}  \left( F_{(1,0)}'' -2  F_{(0,1)}'' \right)-\frac{1}{720}  F_{(0,0)}^{(6)} \right)+\frac{1}{1152} R^8 (F_{0}^{(4)})^2\Bigg).
\end{aligned}
\ee
By long computations with direct variations from refined modular anomaly equations, we find
\be
\delta\Lambda_{(0,2)}=\frac{1}{2}\Lambda_{(0,1)}''+F_{(0,1)}'\Lambda_{(0,1)}'.
\ee
Similar computations give
\be
\delta \Lambda_{(1,1)}=\frac{1}{2}\Lambda_{(1,0)}''+F_{(1,0)}'\Lambda_{(0,1)}'+F_{(0,1)}'\Lambda_{(1,0)}'\qquad\textrm{and}\qquad \delta \Lambda_{(2,0)}=F_{(1,0)}'\Lambda_{(1,0)}'.
\ee

In summary, we find using refined modular anomaly equations, the $\Lambda_{(n,g)}$ factors have the following modular anomaly
\be
\delta\Lambda_{(n,g)}=\frac{1}{2}\Lambda_{(n,g-1)}''+\sum_{n_1,g_1}F_{(n_1,g_1)}'\Lambda_{(n-n_1,g-g_1)}'
\ee
where the summation is $n_1$ from 0 to $n$, $g_1$ from 0 to $g$, but $(n_1,g_1)\neq (0,0)$ or $(n,g)$. We checked this equation for all $(n,g)$ with $n+g\le 6$.

\subsection{Unity odd orders}
For odd orders, $n\in\IZ+1/2$. The situation is a bit more complicated. The first $\Lambda$ factor is
\be
\Lambda_{(\frac12,0)}=\sum\Theta(R) \left(-\frac{1}{6} R^3 F_{(0,0)}^{(3)}+R\left(F_{(0,1)}'+F_{(1,0)}'\right)\right).
\ee
Recall $f(n):=\sum_R \Theta(t,R)R^{2n}$, thus
\be
\frac{1}{6}F_{(0,0)}^{(3)}f(\frac32)=-\Lambda_{(\frac12,0)}+\left(F_{(0,1)}'+F_{(1,0)}'\right) f(\frac12).
\ee
The first contact term equation means $\Lambda_{(\frac12,0)}'=0$ and $\delta\Lambda_{(\frac12,0)}=0$. From the above $f(\frac32)$ and recursive relation (\ref{fnrecur}), one can compute all $f(n)$ for $n=\frac52,\frac72,\frac92,\dots$ 

The subleading order is $n+g=\frac32$. From the expansion of blowup equations, we find
\be\nonumber
\begin{aligned}
\Lambda_{(\frac12,1)}=&\sum\Theta(R)\Bigg(R \left(4 F_{(0,2)}'+(3 F_{(0,2)}-2 F_{(1,1)}+F_{(2,0)}) \left(F_{(0,1)}'+F_{(1,0)}'\right)+F_{(1,1)}'-2 F_{(2,0)}'\right)\\
&-\frac{1}{6} R^3 \left(3 \left(F_{(0,1)}'+F_{(1,0)}'\right) \left(2 F_{(0,1)}''-F_{(1,0)}''\right)+(3 F_{(0,2)}-2 F_{(1,1)}+F_{(2,0)}) F_{(0,0)}^{(3)}+3 F_{(0,1)}^{(3)}\right)\\
&+R^5 \left(\frac{1}{24} \left(F_{(0,0)}^{(4)} \left(F_{(0,1)}'+F_{(1,0)}'\right)+2 F_{(0,0)}^{(3)} \left(2 F_{(0,1)}''-F_{(1,0)}''\right)\right)+\frac{1}{60} F_{(0,0)}^{(5)}\right)\\
&-\frac{1}{144} R^7 F_{(0,0)}^{(3)} F_{(0,0)}^{(4)}\Bigg),
\end{aligned}
\ee
and
\be\nonumber
\begin{aligned}
\Lambda_{(\frac32,0)}=&\sum\Theta(R)\Bigg(R \left(-F_{(0,2)}'-(F_{(0,2)}+3 F_{(2,0)}) \left(F_{(0,1)}'+F_{(1,0)}'\right)+F_{(2,0)}'\right)\\
&+\frac{1}{6} R^3 \left(3 \left(F_{(0,1)}'+F_{(1,0)}'\right) F_{(0,1)}''+\left(F_{(0,1)}'+F_{(1,0)}'\right)^3+(F_{(0,2)}+3 F_{(2,0)}) F_{(0,0)}^{(3)}+F_{(0,1)}^{(3)}\right)\\
&+\frac{1}{120} R^5 \left(-10 F_{(0,0)}^{(3)} \left(\left(F_{(0,1)}'+F_{(1,0)}'\right)^2+F_{(0,1)}''\right)-5 F_{(0,0)}^{(4)} \left(F_{(0,1)}'+F_{(1,0)}'\right)-F_{(0,0)}^{(5)}\right)\\
&+\frac{1}{144} R^7 F_{(0,0)}^{(3)} \left(2 F_{(0,0)}^{(3)} \left(F_{(0,1)}'+F_{(1,0)}'\right)+F_{(0,0)}^{(4)}\right)-\frac{1}{1296}R^9 \left(F_{(0,0)}^{(3)}\right)^3\Bigg).
\end{aligned}
\ee
To compute their modular anomaly, we need first replace all $\sum \Theta(t,R) R^{2n+1}$ here to $\sum \Theta(t,R) R$ and $\Lambda_{(\frac12,0)}$. For example, for $(n,g)=(\frac12,1)$, after the replacements we find
\be
\begin{aligned}
\Lambda_{(\frac12,1)}=A\Lambda_{(\frac12,0)}+B f(\frac12),
\end{aligned}
\ee
where
\be\nonumber
\begin{aligned}
A=&\,3 F_{(0,2)}-2 F_{(1,1)}+F_{(2,0)}+\left(F_{(0,0)}^{(3)}\right)^{-1}\left(\left(F_{(1,0)}'-F_{(0,1)}'\right) \left(2 F_{(0,1)}''-F_{(1,0)}''\right)+3 F_{(0,1)}^{(3)}\right)\\
&+\left(F_{(0,0)}^{(3)}\right)^{-2}\left(\frac{1}{6} F_{(0,0)}^{(4)} \left(\left(F_{(0,1)}'-F_{(1,0)}'\right)^2-19 F_{(0,1)}''+F_{(1,0)}''\right)-\frac{2}{5} F_{(0,0)}^{(5)} \left(2 F_{(0,1)}'+F_{(1,0)}'\right)\right)\\ 
&+\left(F_{(0,0)}^{(3)}\right)^{-3}\left( \frac{1}{30} F_{(0,0)}^{(4)} \left(15 F_{(0,0)}^{(4)} \left(3 F_{(0,1)}'+F_{(1,0)}'\right)-11 F_{(0,0)}^{(5)}\right)  \right)+\frac{1}{2}\left(F_{(0,0)}^{(3)}\right)^{-4} \left(F_{(0,0)}^{(4)}\right)^3
,
\end{aligned}
\ee
and
\be\nonumber
\begin{aligned}
B=&\, 4 F_{(0,2)}'+F_{(1,1)}'-2 F_{(2,0)}'+\left(F_{(0,0)}^{(3)}\right)^{-1}\bigg(-3 F_{(0,1)}^{(3)} \left(F_{(0,1)}'+F_{(1,0)}'\right)-2 \left(F_{(0,1)}''\right)^2-F_{(1,0)}'' F_{(0,1)}''\\
&+\left(F_{(1,0)}''\right)^2\bigg)
+\left(F_{(0,0)}^{(3)}\right)^{-2}\bigg(\frac{1}{6} F_{(0,0)}^{(4)} \left(18 \left(F_{(0,1)}'+F_{(1,0)}'\right) F_{(0,1)}''-F_{(0,1)}^{(3)}-F_{(1,0)}^{(3)}\right)\\
&+\frac{1}{5} F_{(0,0)}^{(5)} \left(3 \left(F_{(0,1)}'+F_{(1,0)}'\right)^2-F_{(0,1)}''-F_{(1,0)}''\right) \bigg)+\frac{1}{30}\left(F_{(0,0)}^{(3)}\right)^{-3} F_{(0,0)}^{(4)} \Big(11 F_{(0,0)}^{(5)} (F_{(0,1)}'+F_{(1,0)}')\\
&-15 F_{(0,0)}^{(4)} (2 (F_{(0,1)}'+F_{(1,0)}')^2-F_{(0,1)}''-F_{(1,0)}'')\Big) -\frac{1}{2}\left(F_{(0,0)}^{(3)}\right)^{-4} \left(F_{(0,0)}^{(4)}\right)^3 (F_{(0,1)}'+F_{(1,0)}')
.
\end{aligned}
\ee
By long computations with variations from refined modular anomaly equations, we find $\delta A=0$ and $\delta B =0$. Thus
\be
\delta\Lambda_{(\frac12,1)} =0.
\ee
Similarly, we compute and find
\be
\delta\Lambda_{(\frac32,0)} =0
\ee
For higher order $\Lambda$ factors, the variation by refined modular anomaly equations does not directly gives zero. Nevertheless, we still find closed formulas for them. For example for $n+g=5/2$, we find
\be
\begin{aligned}
\delta\Lambda_{(\frac12,2)} &=\frac12\Lambda_{(\frac12,1)}''+F_{(0,1)}'\Lambda_{(\frac12,1)}'\ ,\\
\delta\Lambda_{(\frac32,1)} &=\frac12\Lambda_{(\frac32,0)}''+F_{(0,1)}'\Lambda_{(\frac32,0)}'+F_{(1,0)}'\Lambda_{(\frac12,1)}'\ ,\\[+2mm]
\delta\Lambda_{(\frac52,0)} &=F_{(1,0)}'\Lambda_{(\frac32,0)}'.
\end{aligned}
\ee
In summary, for $n\in \mathbb{Z}+1/2$ and $g\in \mathbb{Z}$, we find
\be
\delta \Lambda_{(n,g)}=\frac{1}{2}\Lambda_{(n,g-1)}''+\sum_{n_1,g_1}F_{(n_1,g_1)}'\Lambda_{(n-n_1,g-g_1)}'\ ,
\ee
where the summation is $n_1$ from 0 to $n-1/2$, $g_1$ from 0 to $g$, but $(n_1,g_1)\neq (0,0)$ or $(n-1/2,g)$. We checked this equation for all $(n,g)$ with $n+g\le 11/2$.

\subsection{Vanishing}
The zeroth contact term equation for the vanishing case is trivial.
 The first contact term equation is
\be
0=\sum_{ N \in \mathbb{ Z}} (-)^{|N|} \left(-\frac{1}{6} R^3 F_{(0,0)}^{(3)}+R\left(F_{(0,1)}'+F_{(1,0)}'\right)\right)\exp{ \left(-\frac{1}{2} R^2 F_{(0,0)}''+F_{(0,1)}-F_{(1,0)}\right)} .
\ee
Noticing $\exp(F_{(0,1)}-F_{(1,0)})$ can be factored out, the above equation is actually integrable:
\be
C=\sum_{ N \in \mathbb{ Z}} (-)^{|N|} R\exp{ \left(-\frac{1}{2} R^2 F_{(0,0)}''+3(F_{(0,1)}+F_{(1,0)})\right)} .
\ee
Here $C$ is a weight one form and independent from $t$. For local $\mathbb{P}^1\times \mathbb{P}^1$ i.e. 5d $\mathcal{N}=1$ $SU(2)$ gauge theory, this was already obtained in the section 5.2 of \cite{Nakajima:2005fg}. For higher orders $n+g\ge 1$, we do not find such integrable structure for the vanishing component equations.


\section{Non-Holomorphic Blowup Equations}\label{sec:nonholo}
The existing literature on blowup equations only involves the holomorphic limit of the refined partition functions for both gauge theory and topological strings. It is natural to ask whether the full refined partition functions with the non-holomorphic dependence still satisfy certain generalized blowup equations. The process to check the consistency equations in previous section indeed suggests a natural non-holomorphic extention of blowup equations.

It is more convenient to discuss the non-holomorphic version of blowup equations in terms of component equations than the original form.
For example, we have the following non-holomorphic component equation for $\Lambda_{(0,1)}$:
\be\nonumber
\begin{aligned}
\Lambda_{(0,1)}(t,\bar{t},r)\!=&\!\sum(-)^{|N|} \Big(3 F_{(0,2)}(t,\bar{t})-2 F_{(1,1)}(t,\bar{t})+F_{(2,0)}(t,\bar{t})-R^2 F''_{(0,1)}(t,\bar{t})+\frac{1}{2}R^2 F''_{(1,0)}(t,\bar{t})\\
&\qquad\qquad+\frac{1}{24} R^4 F_{0}^{(4)}(t,\bar{t})\Big)\exp{ \left(-\frac{1}{2} R^2 F_{(0,0)}''(t)+F_{(0,1)}(t)-F_{(1,0)}(t)\right)}.
\end{aligned}
\ee
Note it is crucial that the exponential part $\Theta(R)$ still only depends on $t$, but not on $\bar{t}$. For an arbitrary $r$ vector, $\Lambda_{(0,1)}(t,\bar{t},r)$ is an almost-holomorphic form of weight zero. However for the $r$ fields, $\Lambda_{(0,1)}(t,\bar{t},r)$ becomes holomorphic and only depends on the mass parameters that is $\Lambda_{(0,1)}(m,r)$.

We can also make the blowup formula compact if we formally define
\begin{multline}
F(t+\epsilon,\bar{t},\eq,\et):=\frac{1}{\eq\et}\(\sum_{m=0}^3\frac{\epsilon^m}{m!}F_{(0,0)}^{(m)}(t)+\sum_{m=4}^\infty\frac{\epsilon^m}{m!}F_{(0,0)}^{(m)}(t,\bar{t})\)\\
+F_{(0,1)}(t)+\sum_{m=1}^\infty\frac{\epsilon^m}{m!}F_{(0,1)}^{(m)}(t,\bar{t})+\frac{(\eq+\et)^2}{\eq\et}
\(F_{(1,0)}(t)+\sum_{m=1}^\infty\frac{\epsilon^m}{m!}F_{(1,0)}^{(m)}(t,\bar{t})\)\\
+\sum_{n+g\ge 2}^{\infty}(\epsilon_1+\epsilon_2)^{2n}(\epsilon_1\epsilon_2)^{g-1}\sum_{m=0}^\infty\frac{\epsilon^m}{m!}F_{(n,g)}^{(m)}(t,\bar{t}).\phantom{-------}
\end{multline}
Then the non-holomorphic version of blowup equations can be written as
\be\label{eq:blowupZnonholo}
\Lambda(t,\bar{t},\eq,\et,r)=	\sum_{ N \in \mathbb{ Z}^{\gS}}  (-)^{|N|}\frac{{Z}(t+ \eq R,\bar{t},\eq,\et-\eq)\,{Z}(t+ \et R,\bar{t},\eq-\et,\et  )}{Z(t,\bar{t},\eq,\et)}.
\ee
For an arbitrary $r$ vector, $\Lambda(t,\bar{t},\eq,\et,r)$ is almost holomorphic. But for the $r$ fields, 
all non-holomorphic dependence cancels in the end such that $\Lambda(t,\bar{t},\eq,\et,r)$ becomes holomorphic and can be written as $\Lambda(m,\eq,\et,r)$.

\section{Summary and open questions}\label{sec:open}
In this short note, we discuss the nontrivial relation between blowup equations and refined holomorphic/modular anomaly equations. We find the latter computes the modular anomaly of the $\Lambda_{(n,g)}$ factors of blowup equations in a recursive way and obtain an elegant closed formula. Such formula governs the consistency between the two rather different type of equations for the free energy of refined topological strings. It also inspires us to propose an non-holomorphic version of blowup equations which realizes the traditional blowup equations as a limit.

Throughout the note, we limit ourselves to refined topological strings on non-compact Calabi-Yau threefolds.
As holomorphic anomaly equations are most powerful for compact Calabi-Yau threefolds (for example quintic \cite{Huang:2006hq}), 
it is intriguing to consider whether blowup equations can be generalized in certain sense to the compact cases. To begin with, one needs to make sense of the refinement for topological strings on compact Calabi-Yau. There is already some progress on the refinement for some cases with elliptic fibration \cite{Huang:2020dbh}. We point out that if blowup equations can be generalized to compact Calabi-Yau threefolds, the proper definition of the $\Lambda$ factor should be anomaly free $\delta\Lambda=0$ since there is no more concept of mass parameters.

Recently blowup equations are also generalized to supersymmetric gauge theories with defects \cite{Nekrasov:2020qcq,Jeong:2020uxz} and Wilson loops \cite{Kim:2021gyj,Chen:2021rek}. In the setting of local Calabi-Yau threefolds, it is natural to consider the open topological string theory.
For open refined topological string with $\eq$-brane, the wave function ${\Psi}_1(t,x,\eq,\et)$ was beautifully studied in \cite{Aganagic:2011mi} where there is one more parameter $x$ labeling the position of the brane on the mirror curve, and $\eq,\et$ are no longer symmetric. The form of generalized blowup equations in \cite{Nekrasov:2020qcq,Jeong:2020uxz,Kim:2021gyj,Chen:2021rek} inspires the following blowup equations for the brane wave function
\be\label{eq:blowupPhi}
\Lambda(t,x,\eq,\et,r)=	\sum_{ N \in \mathbb{ Z}^{\gS}}  (-)^{|N|}\frac{{\Psi}_1(t+ \eq R,x,\eq,\et-\eq)\,{Z}(t+ \et R,\eq-\et,\et  )}{\Psi_1(t,x,\eq,\et)}.
\ee
It will be very interesting to clarify this equation and its relation with open holomorphic anomaly equations \cite{Walcher:2007tp,Cook:2007dj,Bonelli:2007gv}.

\section*{Acknowledgements}

This work was finished in 2020 summer when the author was in Max Planck Institute for Mathematics in Bonn. The author would like to thank Min-xin Huang, Albrecht Klemm, Haowu Wang, Xin Wang and Don Zagier for useful discussion. 
The main question in this paper was raised when the author worked on \cite{Huang:2017mis} with Min-xin Huang and Xin Wang in late 2016. 
The form of the consistency equation has appeared in the author's PhD thesis in SISSA \cite{Sun:2020jud}.
The author is currently supported by KIAS Grant QP081001.

\end{document}